\documentclass[usenatbib]{aa} 

\usepackage{epsfig} 
\usepackage{graphicx} 
\usepackage{natbib}  
\def\pl{{\sc pl}} 
 
\def\compps{{\sc compps}}

\def\bb{{\sc bb}}

\def\gs{{\sc Gaussian}} 
\def\nh{{$N_{\rm H}$}} 
\def\dbb{{\sc diskbb}} 
\def\bb{{\sc bb}}

\def\hete{HETE~J1900.1--2455} 
\def\be{\begin{equation}} 
\def\ee{\end{equation}} 
 
\begin{document} 
   \title{Simultaneous INTEGRAL and RXTE observations of the accreting 
   millisecond pulsar HETE~J1900.1--2455} 
 
 \author{M. Falanga\inst{1,2}\fnmsep\thanks{\email{mfalanga@cea.fr}} 
\and  J. Poutanen\inst{3}
\and E. W. Bonning\inst{4} 
\and L. Kuiper\inst{5} 
\and J. M. Bonnet-Bidaud\inst{1} 
\and A. Goldwurm\inst{1,2}  
\and W. Hermsen\inst{5,6} 
\and L. Stella\inst{7}           
          } 
 
\offprints{M. Falanga} 
\titlerunning{INTEGRAL/RXTE observations of HETE~J1900.1-2455}  
\authorrunning{M. Falanga et al.}  
  
\institute{CEA Saclay, DSM/DAPNIA/Service d'Astrophysique (CNRS FRE  
  2591), F-91191, Gif sur Yvette, France 
\and Unit\'e mixte de recherche Astroparticule et  
Cosmologie, 11 place Berthelot, 75005 Paris, France  
\and  Astronomy Division, P.O.Box 3000, FIN-90014 University of  
  Oulu, Finland  
\and Laboratoire de l'Univers et de ses Th'{e}ories, Observatoire de  
Paris, F-92195 Meudon Cedex, France  
\and SRON Netherlands Institute for Space Research, Sorbonnelaan 2, 
3584 CA Utrecht, The Netherlands 
\and Astronomical Institute ``Anton Pannekoek'', University of 
Amsterdam, Kruislaan 403, NL-1098 SJ Amsterdam, The Netherlands 
\and INAF-Osservatorio Astronomico di Roma, via Frascati 33, 00040  
  Monteporzio Catone (Roma), Italy  
             } 
 
   \date{ } 
 
  
  \abstract 
{} 
   {\hete\ is the seventh known  X-ray transient accreting
  millisecond pulsar and has been in outburst for more than one year. 
  We compared the data on \hete\ with other similar objects 
  and made an attempt at deriving constraints on the physical
  processes responsible for a spectral formation.}
   {The broad-band spectrum of the persistent emission 
  in the 2--300 keV energy band and the timing
  properties were studied using simultaneous {{\it INTEGRAL} } and
  publicly available 
  {{\it RXTE}} data obtained in October 2005. 
  The properties  of the X-ray bursts observed from \hete\ 
  were also investigated.}   
   {The spectrum is well described by a two-component model consisting of
  a blackbody-like soft X-ray emission at 0.8 keV temperature and  a thermal
  Comptonized spectrum with electron temperature of 30 keV and Thomson
  optical depth $\tau_{\rm T} \sim 2$ for the slab geometry. The
  source is detected by {{\it INTEGRAL}} up to 200 keV at a luminosity
 of $5\times10^{36}$ erg s$^{-1}$ (assuming a distance of 5 kpc)
  in the 0.1--200 keV energy band. We have also detected one type I
  X-ray burst which shows  photospheric radius expansion. The burst
  occurred at an inferred persistent emission level of $\sim$ 3--4\%
  of the Eddington luminosity. Using  data for all
  X-ray bursts observed  
  to date from \hete, the burst recurrence time is estimated to be
  about 2 days. 
  No pulsations  have been detected either in the {{\it RXTE}}
  or in the {{\it INTEGRAL}} data which puts interesting constraints on 
  theories of  magnetic field evolution in neutron star
  low-mass X-ray binaries. 
  }
   {} 
 
 
   \maketitle

\section{Introduction} 
\label{sec:intro} 
 
The detection of  X-ray millisecond pulsation in persistent emission 
from low-mass X-ray binaries (LMXBs) 
remained elusive for many years until the discovery of the first 
accreting millisecond pulsar (MSP) by \citet{wvdk98}. Since that time, a 
total of seven accreting MSP transients  have been detected. They are weakly 
magnetized ($\sim 10^{8}-10^9$ G) neutron stars (NS) with spin 
frequencies in the 180--600 Hz range and   
orbital periods between 40 min and 5 hr \citep[see reviews by][]{w05,p06}. 
Their companion stars have been found to be either highly   evolved  white 
  or brown dwarfs. For the first time,  the predicted 
decrease of the NS spin period during accretion was measured  in the
accreting MSP IGR~J00291+5934 \citep{mfb05,burd06}. This provided a strong
confirmation of the theory of `recycled' pulsars in which old
neutron stars in LMXBs become millisecond radio pulsars through
spin-up by transfer of angular momentum by the accreting material.  
 
MSP energy spectra can be well  described by a two-component model
  consisting of a soft black body (or multi-color  blackbody) and
a hard power-law like tail.
The soft thermal component could be associated with  
radiation from the accretion disc and/or the heated NS surface around the 
shock \citep[see e.g.][]{gp05,p06}. The hard emission is likely to be
produced by thermal Comptonization in the hot accretion shock on the
NS surface \citep{gdb02,pg03} with seed photons coming from the
stellar surface. The observed hard spectra are similar to the spectra
observed from atoll sources in their hard, low-luminosity state \citep{b00}.

\hete\ was discovered during a bright X-ray burst by the {\it High Energy 
Transient Explorer 2} ({\it HETE-2}) on 14 June 2005 \citep{vanderspek05}. 
Followup observations with the {\it Rossi X-ray Timing Explorer} ({\it RXTE}) 
identified the source as the seventh X-ray accreting millisecond   
pulsar, with a pulse frequency of 377.3 Hz, an orbital period  
of 83 min, and most likely a 0.016--0.07 M$_{\odot}$ brown dwarf companion 
\citep{kaaret06}. The detected burst was consistent with a type I 
X-ray burst with  photospheric radius expansion.   
Assuming that the bolometric burst peak luminosity 
during  photospheric radius expansion saturated at the Eddington 
limit, \citet{kawai05} estimated the distance to the source to be
$\sim5$ kpc assuming helium burst burning and canonical NS values.   
 
An optical counterpart candidate had an $R$-band 
magnitude of 18.02 and a broad HeII emission line spectrum 
\citep[][]{fox05,steeghs05a}. A similar line was previously 
observed in  IGR~J00291+5934 \citep{rjs04,ffc04}. 
The optical counterpart of the X-ray source is located at 
coordinates $\alpha_{\rm J2000} = 19^{\rm h}00^{\rm m}08\fs65$ and
$\delta_{\rm J2000} = -24{\degr}55\arcmin13\farcs7$ with an
uncertainty of $0\farcs2$. Near-infrared observations detected the optical
candidate at a constant  magnitude of J=17.6. No radio
counterpart consistent with the \hete\ coordinates was  detected by
the VLA \citep[][]{steeghs05b,rupen05}.  
 
\begin{figure} 
\centerline{\epsfig{file=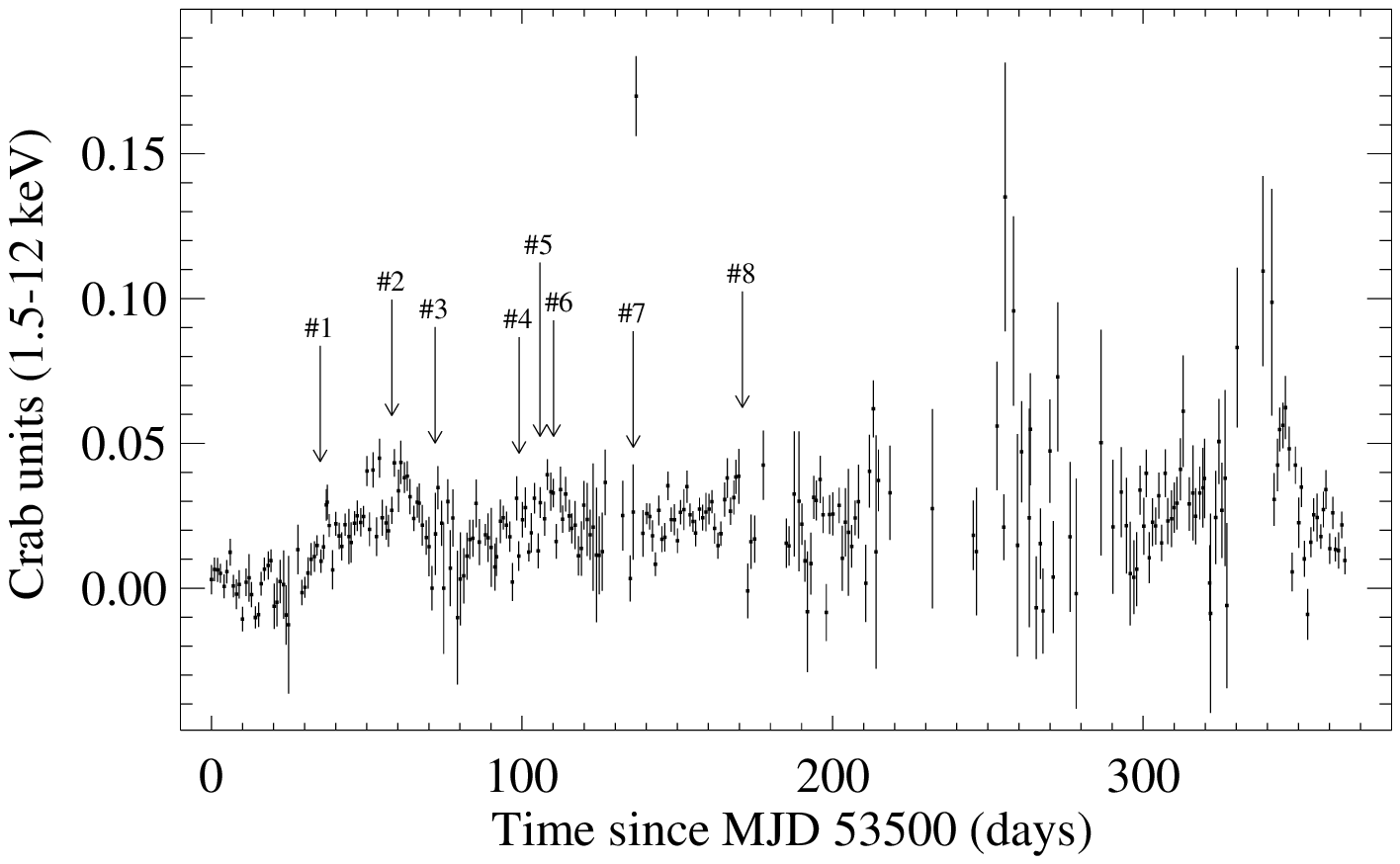,width=8.0cm} } 
\caption{{\it RXTE}/ASM light curve for \hete\ averaged over 1-day 
intervals from May 5, 2005 (53500 MJD).  
The count rate has been converted into flux 
using 1 Crab Unit for 75 cts s$^{-1}$ \citep{l96}. The arrows indicate
the times of the detected X-ray bursts \citep[][and this
  work]{vanderspek05,barbier05,galloway05}.  
} 
\label{fig:asm} 
\end{figure}

One year after the discovery, \hete\ is still active (see Fig. \ref{fig:asm}). 
Compared to  other accreting MSPs with outburst periods of a few
days to a month, \hete\ shows evidence of being a ``quasi-persistent''
X-ray source. This source also has  other properties atypical for accreting
MSPs. During the first $\sim30$ days after the discovery, \hete\ showed
significant flux emission variability in the fractional rms amplitude
\citep{galloway06b}.  
On July 8, 2005 (MJD 53559) during the flux  
brightening, the observed pulse frequency decreased by
$\Delta\nu/\nu\sim6\times10^{-7}$, and in the subsequent observations,
the pulsations were suppressed  \citep{kaaret06}.  
 
In this paper we report the  {\it INTEGRAL} observations of \hete\ obtained 
simultaneously with {\it RXTE}.  
We study the broad-band spectral and timing properties of the 
source. The X-ray burst properties are also investigated.

\section{Observations and data} 
 
\subsection{INTEGRAL} 
\label{sec:integral}  
 
The present data were obtained during the 
{\em INTEGRAL}  \citep{w03} Target of Opportunity  (ToO) observation  during 
satellite revolution 371, starting on October 27 and ending on October
29, 2005,  
with a total exposure time of 210 ks. 
The observation, aimed at  \hete, consists of 64 stable 
pointings with a source position offset $<2\fdg1$.   
We analyzed data from the IBIS/ISGRI coded mask telescope \citep{u03,lebr03} 
at energies between 18 and 300 keV and from the JEM-X monitor, module 1 
\citep{lund03} between 3 and 20 keV.   
The data reduction was performed using the standard Offline Science  
Analysis (OSA) version 5.1 distributed by the {\it INTEGRAL} Science 
Data Center \citep{c03}. The algorithms used for the spatial and 
spectral analysis are described in \citet{gold03}. The ISGRI light 
curves are based on events selected according to the detector  
illumination pattern  for \hete. For ISGRI we used an illumination 
factor threshold of 0.25 for the energy range 18--40 keV; for JEM-X we 
used the event list of the whole detector in the 3--6 keV, 6--12 keV 
and 12--20 keV energy band.

We first deconvolved and analyzed separately the 64 single pointings and 
then combined them into a total mosaic image in the 20--40 keV and 
40--200 keV energy band, respectively. In the mosaic, \hete\ is
clearly detected at a significance level of $\sim100\sigma$ (20--40 keV) and 
$\sim50\sigma$ at higher energy (40--200 keV). 
The source position in the 20--40 keV band is
$\alpha_{\rm J2000} = 19^{\rm h}00^{\rm m}09\fs19$ and $\delta_{\rm J2000} = 
-24{\degr}55\arcmin12\farcs1$ (error of $0\farcm4$ at the 
90 per cent confidence  level, \citealt{gros03}), which 
is offset with respect to the optical position by
$0\farcm13$  \citep{fox05}. 
The background-subtracted 20--40 and 40--200 keV light curves were 
extracted from the images using all available pointings, each with 
$\sim 3.3$ ks exposure. The source mean count rate was almost constant 
at $\sim 4.0$ cts s$^{-1}$ ($\sim 3.0\times10^{-10}$ erg cm$^{-2}$ 
s$^{-1}$) in the 20--40 keV band  and $\sim 2.8$ cts s$^{-1}$ ($\sim 
3.8\times10^{-10}$ erg cm$^{-2}$ s$^{-1}$) in the 40--200 keV energy band. 
The count rates were converted to un-absorbed flux using 
the Comptonization model described in Sect. \ref{sec:spec_persistent}.

\subsection{RXTE} 
\label{sec:rxte}  
 
We used two {\it RXTE} ToO observations performed simultaneously with 
{\it INTEGRAL} between October 28 and 29, 2005 (observation 
id. 91432), for which the data were publicly available. The total  
net exposure times were 6.2 and 3.2 ks, respectively.    
We analyzed data from the Proportional Counter Array 
(PCA; 2--60 keV) \citep{jahoda96} and the High Energy X-ray Timing  
Experiment (HEXTE; 15--250 keV) \citep{rothschild98} on board the {\it RXTE} 
satellite. For the spectral analysis, we extracted the PCA (modules 0, 
2 and 3) energy spectrum using the standard software package FTOOLS 
version 6.0.2.  
For HEXTE, we used the ON-source data, using default screening criteria for 
Cluster 0.  
For the timing analysis we used all PCU data, which were on 
during the entire observation. The PCA data were collected in the {\tt 
  E\_125us\_64M\_0\_1s} event mode, recording event arrival times with 
125 $\mu$s time resolution, and sorting events in 64 PHA 
channels. Default selection criteria were applied.

\subsection{Spectral analysis} 
 
Broad-band spectral analysis was done using XSPEC version 11.3 
\citep{arnaud96}, combining the 3--22 keV {\it RXTE}/PCA data with 
the 3--22 keV {\it INTEGRAL}/JEM-X and 16--90 keV {\it RXTE}/HEXTE data with 
20--300 keV {\it INTEGRAL}/ISGRI data. A multiplicative factor for each 
instrument was included in the fit 
to take into account the uncertainty in the cross-calibration of the 
instruments. The factor was fixed at 1 for the PCA data. A systematic 
error of 1\% was applied to PCA/HEXTE and  2\% to JEM-X/ISGRI spectra, 
which corresponds to the current uncertainties in the response 
matrix. All uncertainties in the spectral parameters are given at a 90\% 
confidence level for single parameters. We use a source distance of 5 
kpc throughout the paper.

\section{Results}

\subsection{Persistent emission} 
 
\subsubsection{Spectral properties} 
\label{sec:spec_persistent} 
 
We analyzed the JEM-X/ISGRI spectra independently both 
before and after the X-ray burst, and no significant spectral variation was 
found. The burst occurred around the middle of the {\it INTEGRAL} observation. 
The {\it RXTE} observations were also performed  before and after the
X-ray burst, and the PCA/HEXTE spectral parameters were consistent with those
determined from the {\it INTEGRAL} spectra.
Therefore, we studied in
detail the broad-band 3--300 keV spectrum of \hete\ using the joint
{\it INTEGRAL} and {\it RXTE} data.  For the {\it INTEGRAL} data we
removed the time interval corresponding to the burst. The energy range
covered by JEM-X/PCA does not allow us to constrain  the
interstellar hydrogen column density, {\nh}, well.  
Therefore, in all our spectral fits we fixed \nh\  at 
$1.6\times10^{21}$ cm$^{-2}$, the value found from {\em Swift} observations 
at lower energies \citep{campana05}. This value is close to 
the Galactic value reported in the radio maps of \citet{dickey90}.  
 
We first fit the joint JEM-X/ISGRI/PCA/HEXTE (3--300 keV) spectrum 
using a simple photoelectrically-absorbed power-law, \pl, model which 
was found inadequate with a $\chi^{2}{\rm /d.o.f.} = 662/110$.  
A better fit was found by adding a blackbody, \bb, model for the soft thermal 
emission and by replacing the \pl\ with a cutoff \pl\ model. 
This gave a $\chi^{2}{\rm /d.o.f.} = 131/107$. The best-fit parameters 
are:  blackbody temperature $kT_{\rm bb}=0.90\pm0.03$ 
keV, a power-law photon index $\Gamma\sim1.60\pm0.03$, 
 and the cutoff energy $\sim59_{-2}^{+3}$ keV.  The PCA data show
 residuals in the  
6--10 keV range that can be fit by a broad Gaussian 
emission line. Correspondingly, the fit improved with a $\chi^{2}{\rm
  /d.o.f.} =  
105/105$. Confining the centroid of the line in the 6.3--6.8 
keV range gave a line width of $\sigma =0.44\pm0.3$ (equivalent width 
$\simeq102$ eV); fixing the line width at 0.5 keV constrains its 
position at $\simeq6.41$ keV.  
 
\begin{figure} 
\includegraphics[angle=-90,width=8.5cm]{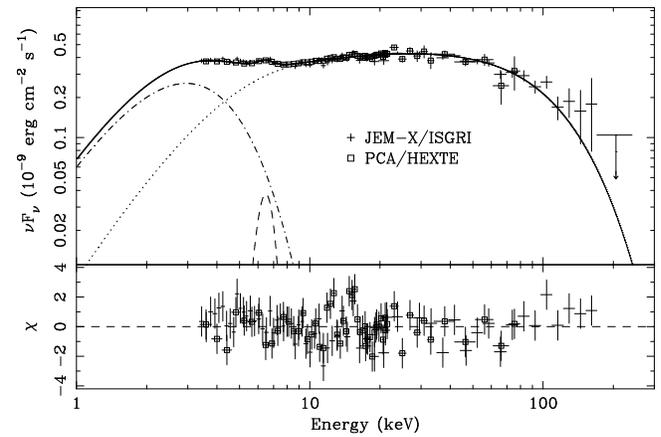} 
\caption{The unfolded spectrum of \hete\ fit with a \compps\ model plus  
a \bb\ and a Gaussian line. The data points correspond 
to the PCA (3--22 keV), JEM-X (3--22 keV), HEXTE (16--90 keV) and 
the ISGRI (20--300 keV) spectra, respectively. The blackbody model 
is shown by a  dot-dashed curve, the dotted curve gives the \compps\ 
model, the  dashed curve is the Gaussian line, and the total spectrum 
is shown by the solid curve. The lower panel shows the residuals 
between the data and the model.
}  
\label{fig:spec} 
\end{figure}

In order to compare the \hete\ spectrum with previously observed spectra 
of the same source class \citep{gdb02,gp05,mfa05,mfb05}, we replaced the 
simple cutoff \pl\ model with the thermal Comptonization model
\compps\ in the slab geometry 
\citep{ps96}. The main model parameters are the Thomson optical depth 
$\tau_{\rm T}$ across the slab, the electron temperature 
$kT_{\rm  e}$, the soft seed photon temperature $kT_{\rm seed}$, and  
the inclination angle $\theta$ between the slab normal and the line of sight. 
The seed photons are assumed to be injected from the bottom of the
slab. The soft thermal emission is fit by a simple blackbody \bb\ or a 
multi-temperature disc blackbody \dbb\ model \citep{mitsuda84}.  
The best fit parameters for the  different models are reported in Table 
\ref{table:spec}. In Fig. \ref{fig:spec}, we show the unfolded spectrum  
and the residuals of the data to the \bb\ plus \compps\ model.

The temperature of the soft thermal emission was found 
to be a factor $\sim2$ higher than in other accreting millisecond pulsars 
\citep[e.g.,][]{gp05,mfa05}. Since its discovery, this source has
shown  a rather high blackbody temperature  \citep{kaaret06}. 
It was suggested for the millisecond pulsar XTE~J1751$-$305 that the 
observed soft component could also be the source of the seed photons 
\citep{gp05}, therefore, we repeated the fit with $kT_{\rm 
  soft}=kT_{\rm seed}$  and obtained only a marginally worse 
$\chi^{2}{\rm  /d.o.f.} = 117/106$. As discussed  by 
\citet{gp05} and \citet{kaaret06},  the true spectrum may consist of
two black-bodies (disk and the heated surface of the NS).   
In adding another black-body component, we found that the new model
does not improve  the fit much.   
As the PCA/JEM-X energy range begins at 3 keV, it is
impossible to search for additional soft emission in these data.

\begin{table}[htb] 
\caption{\label{table:spec}Best-fit spectral parameters with  
\gs\ + \compps\ (or cutoff \pl) + \bb\ (or \dbb) model.}   
\begin{tabular}{lllll} 
\hline 
&  \pl\ + \bb\ & {\sc diskbb} & \bb\ & \bb\ \\ 
\hline 
\noalign{\smallskip}  
$N_{\rm H} (10^{22} {\rm cm}^{-2})$ & $0.16$ (f)  & $0.16$ (f) & $0.16$ (f) & $0.16$ (f)\\ 
$kT_{\rm in/bb}$ (keV) & $0.90^{+0.03}_{-0.03}$  & 1.0$^{+0.05}_{-0.03}$ & 
0.8$^{+0.02}_{-0.02}$ & 0.7$^{+0.01}_{-0.01}$\\ 
$R_{\rm in}^{\ a} \sqrt{\cos i}$ (km)  &  -- & $2.6^{+0.2}_{-0.3}$ & -- & --\\ 
$R_{\rm bb}^{\ a}$ (km)  &  $5.0^{+0.7}_{-0.6}$  & -- & $4.8^{+0.7}_{-0.6}$ & $4.4^{+0.6}_{-0.5}$\\ 
$E_{\rm Fe}$ (keV)& $6.41^{+0.13}_{-0.14}$ & 6.41$_{-0.12}^{+0.14}$ & 6.42$_{-0.15}^{+0.1}$ & 6.41$_{-0.14}^{+0.12}$ \\ 
$\sigma_{\rm Fe}$ (keV)& $0.5$ (f) & 0.5 (f) & 0.5 (f) & 0.5 (f) \\ 
$kT_{\rm e}$ (keV)&   --    & $27.2^{+1.2}_{-2.0}$ & $27.9^{+1.8}_{-1.4}$ & $25.9^{+1.9}_{-1.6}$\\ 
$kT_{\rm seed}$ (keV)&  --  & $1.4^{+0.16}_{-0.32}$ & 1.4 (f) & $=kT_{\rm bb}$\\ 
$\tau_{\rm T}$   &    --    & $2.1^{+0.1}_{-0.07}$ & $2.0^{+0.06}_{-0.1}$ & $2.2^{+0.1}_{-0.1}$\\ 
$A_{\rm seed}^{a}$ (km$^2$)& -- & $13.8^{+0.3}_{-0.4}$  & 
$14.2^{+0.3}_{-0.3}$& $240^{+42}_{-29}$\\ 
cos $\theta $    &    --    & $0.59^{+0.05}_{-0.07}$  & 0.6 (f)  & 0.6
(f)\\ 
$\Gamma$    &  $1.60^{+0.03}_{-0.03}$     & -- & --  & -- \\ 
$E_{\rm cut}$ (keV)   &  $59^{+3}_{-2}$     & -- & --  & -- \\ 
$\chi^{2}/{\rm dof}$  &  105/105               & 114/103  & 110/105 & 117/106\\ 
$L_{\rm bol}^{a}$($10^{36}$erg/s) & 5.0$\pm$0.5 & 5.6$\pm$0.5 & 4.9$\pm$0.5 & 4.9$\pm$0.5\\ 
\noalign{\smallskip}  
\hline  
\noalign{\smallskip}  
\multicolumn{4}{l}{$^{a}$  Assuming a distance of 5 kpc.} 
\end{tabular}  
\end{table} 
 
\subsubsection{Timing analysis} 
 
We carried out a Z$_{1}^{2}$-statistic \citep{b83} period search of \hete\
using barycentered (JPL DE200 solar system ephemeris)  and orbital
motion \citep{kaaret06} corrected arrival times registered
by the PCA instrument during the two short pointings of the ToO observation 
in the 2--10 keV and 10--20 keV energy bands. 
We centered our periodicity search around the last measured frequency
value 377.2959 
Hz\footnote{Note the frequency quoted in \citet{kaaret06},
377.291596(16) Hz is a typo, the correct value is
377.29596(16) Hz.}, i.e. after the frequency shift observed on 53559 MJD
\citep[see][]{kaaret06}, with a rather wide frequency range of
$\pm0.007$ Hz. Within the scanned frequency range we did not find
any indication of significant pulsations in either of the two energy bands.
Thus, unless there was  another
frequency shift between 53567 and 53670 MJD such that the frequency
range of our search was too small,  there is no
evidence for a pulsed signal from \hete in the two short {\it RXTE}
observations falling within the {\it INTEGRAL} observation.

Thinking that the pulsed fraction might increase above 20--30 keV, 
as was found for IGR~J00291-5934 \citep{mfb05}, we also searched the
high-energy {\it INTEGRAL}  
data for the 2.65 ms period. No coherent pulsed signal was found in
the 18--150 keV  IBIS/ISGRI band near the expected pulse frequency of
\hete, confirming the suppression of the pulsed
signal. \citet{galloway06b} studied the pulsed fraction in detail 
 and find an upper limit of about 1\% rms during our observation. 

\subsection{Properties of the X-ray bursts} 
 
\subsubsection{Burst light curves} 
\label{sec:burst} 
   
\begin{figure} 
\includegraphics[angle=-90,width=8cm]{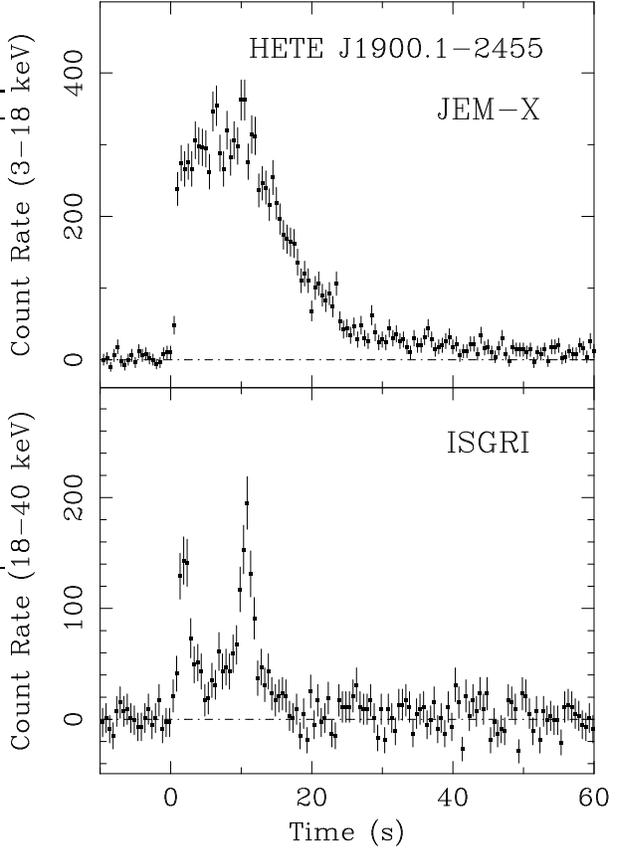} 
\caption{A bright X-ray burst detected from \hete. The JEM-X 
  (3--20 keV, upper panel) and IBIS/ISGRI (18--40 keV; lower panel) 
  net light curves are shown (background subtracted). The time bin is 
  0.5 s for both IBIS/ISGRI and JEM-X light curve. At high energy the 
  burst shows strong evidence of photospheric radius expansion.} 
\label{fig:lc_burst} 
\end{figure}

In Figure \ref{fig:lc_burst} we show the JEM-X and ISGRI burst light 
curve (28 October 2005, 10:25:12 UTC) in different energy bands. The
burst rise time was $0.23\pm0.05$ s. The double peak profile is
clearly evident at  
high energy (lower panel) within the first 12 s, while  
during this time the intensity at lower energy (upper panel) remains
constant.  This can be interpreted as a consequence of a photospheric
radius  expansion (PRE) episode during the  first part of the outburst
\citep[see  e.g.,][]{vh95}. When a burst undergoes a PRE episode, 
the luminosity remains nearly constant at 
the Eddington value, the atmosphere expands, and its temperature  
decreases, resulting in  a double-peak profile  observed at high
energies. The tail of the burst at high energy can be seen for about
5~s after the PRE episode.   
 
\subsubsection{X-ray burst spectra} 
\label{sec:spec_burst} 
 
\begin{figure} 
\includegraphics[angle=-90,width=8cm]{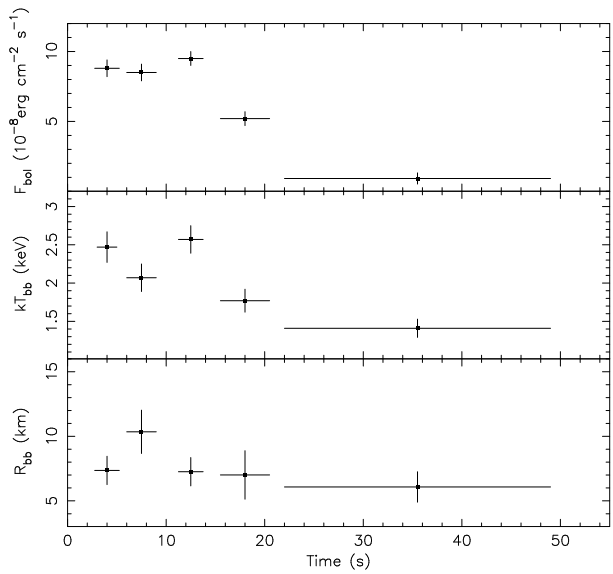} 
\caption{Result of the X-ray burst time-resolved spectral analysis. The 
apparent blackbody radius was estimated assuming a source distance of 5 kpc.} 
\label{fig:burst} 
\end{figure}

We performed time-resolved analysis of the burst spectrum based on the 
JEM-X/ISGRI  3-50 keV energy band data. We verified that during the burst 
pointing the count rate was stable; we then used the JEM-X/ISGRI 
persistent emission spectrum as background. The burst is divided into time 
intervals as shown in Fig. \ref{fig:burst}, the shortest intervals being 
$\sim$ 3 s during the two spikes observed at high energy (see Fig. 
\ref{fig:burst}). The burst net spectrum was well fit by a 
photoelectrically absorbed blackbody ($\chi^{2}_{\rm red} \sim 1.1$). The 
inferred blackbody temperature, $kT_{\rm bb}$, and apparent blackbody  
radius, $R_{\rm bb}$, are shown in the middle and lower panels, 
respectively. During the first 12 s, the un-absorbed 
bolometric flux was almost constant at $F_{\rm peak}=9.5(2)\times10^{-8}$ erg 
cm$^{-2}$ s$^{-1}$, while the blackbody temperature dropped in the middle, 
simultaneous with an increase by a factor of $\sim1.5$ in blackbody 
radius. The observed temperature reached a peak at $\sim$ 2.5 keV, and then 
gradually decreased. The softening of the emission towards the end of the 
decay phase is also indicated by the e-folding decay times 
of $12.5\pm0.5$ s in the 3--6 keV to $4.3\pm0.7$ s in the 12--20 keV
energy band. This behavior is typically observed during PRE X-ray bursts 
\citep[e.g.,][]{kuulkers03}. 
 
The burst fluence, $f_{\rm b}=1.67(6)\times10^{-8}$erg cm$^{-2}$, is
calculated by integrating the measured $F_{\rm bol,bb}$ over the burst
duration of $\sim 50$ s. The effective burst duration is $\tau=f_{\rm b}/F_{\rm
  peak}=18.2(8)$ s, and the ratio of the observed persistent flux to
the net peak flux is $\gamma=F_{\rm pers}/F_{\rm  peak}=0.021(1)$.  
The burst has the same spectral parameters as  previous 
bursts  observed for this source with {\it HETE-2} \citep{kawai05} and
{\it RXTE} \citep{galloway05,galloway06}. 
 Assuming a helium-burst at the Eddington limit \citep{lvpt93} and canonical 
NS parameters (1.4 solar mass and radius of 10 km), we 
estimate the source distance to be $\sim5$ kpc.

\section{Discussion} 
\label{sec:discussion} 
 
We analyzed the {\it INTEGRAL} and {\it RXTE} data of the
accretion-powered millisecond pulsar \hete. We observed the 
source remaining in a bright state $\sim 136$ days after its discovery with 
a bolometric luminosity during our observation of 3--4 per cent
of $L_{\rm Edd}$ \citep{fr02}
(see Fig. \ref{fig:asm} for the {\it RXTE}/ASM 1.5--12 keV light
curve). 
The outburst  light curve of \hete\  is 
completely different from that of the other six observed MSPs
which have shown exponential decays 
during the outburst: declining from  peak luminosities of 3--30 per cent 
$L_{\rm Edd}$ with a decay time-scale of several days up to $\sim 120$ days.  
Only \hete\   has shown a fairly steady behavior at a low 
mass accretion rate (between 1 and 4 per cent of Eddington) with no 
sign of a transition toward quiescence. The most peculiar behavior of \hete\ 
is the suppression of the pulsation at the NS spin period  
about a month from the beginning of the outburst  \citep{kaaret06}; 
all other MSPs have had detectable
pulsations throughout the duration of their outburst. The missing
pulsations and the  persistent X-ray emission at low mass
accretion rate shows \hete\  to  be similar to the most commonly
  observed persistant, non-pulsating LMXB systems. It seems that the
price to pay for being active over a long time is that pulsations are
suppressed, i.e. there is an evident link between the long period of
mass accretion and the missing pulsations.

The best spectral fit to the data required a two-component model, a cutoff \pl\
or a thermal Comptonization model together with a soft component (see 
Table \ref{table:spec}). The soft thermal emission detected at low energies  
could come either from the accretion disk or from a hot spot on the NS surface.
As for the hard component, it  most likely originates from the Comptonization 
of soft seed photons, $kT_{\rm seed}\sim 1.5$ keV in a hot $kT_{\rm e} 
\sim 30$ keV  plasma of moderate optical thickness $\tau_{\rm T} \sim 2$.  
The hard spectral component up to 200 keV contributes most 
of the observed flux (70\%), even though a soft blackbody component is 
needed to fit the data. For the source distance of 5 kpc, the 
unabsorbed 0.1--300 keV luminosity was $(4.9-5.6)\ 10^{36}$erg s$^{-1}$ 
(for different models).

We could not distinguish between the multi-temperature blackbody and 
single blackbody models, as both gave  comparable parameters and $\chi^2$.   
However, for a distance of 5 kpc, our fit for a disk blackbody gives 
an  inner disk radius, $R_{\rm in} \sqrt{\cos\,i} = 2.6$ km, 
smaller than the expected NS radius. 
For blackbody emission, the fit implies an apparent area of the
emission region  
$A_{\rm seed} \sim 14$ km$^2$, which could be consistent  with a
heated NS surface around the accretion shock \citep{gdb02,pg03,gp05}.
On the other hand, the observed spectra before MJD 53558  are
  similar to those observed by us when pulsations are absent \citep{kaaret06}.
 They are also very similar to those of the atoll sources at low  
luminosities (Barret et al., 2000), where the X-rays are probably
produced in the boundary/spreading
layer near the NS equator  \citep{KW91,is99,sp06}.  

Spectral similarities can be explained if in both types of sources
(accreting MSPs and non-pulsating atoll sources), the energy dissipation 
happens in the optically thin medium (i.e. accretion shock and
  boundary/spreading layer)  and the spectral properties 
are determined solely by energy balance and  feedback 
from the NS surface which provides cooling in the form of soft photons
\citep*[see e.g.][]{hm93,stern95,ps96,mbp01}. 
In these two-phase models (cool neutron star with a hot dissipation 
region above), the product of electron temperature and optical depth
is approximately constant.  
In \hete\ we observe $kT_{\rm e}\times \tau_{\rm T}\approx 57$ keV
(see Table 1) which is consistent with the values determined for other MSPs
\citep{pg03,gp05,mfa05,mfb05,p06} as well as with the theoretical models.

Since we found a fairly high blackbody temperature 
compared to other MSP sources, we attempted to fit a blackbody model
with the soft emission being the source of the seed photons, i.e. $kT_{\rm 
  bb}=kT_{\rm seed}$. The obtained parameters are consistent with the 
other models, with similar optical depth, plasma temperature, and 
slightly lower soft thermal emission temperature. 
However, this fit implies a much larger apparent area of the seed photons, 
$A_{\rm seed} \sim 240$ km$^2$, inconsistent  with the whole NS
surface.  

Several X-ray bursts have been observed for this source by various 
observatories. These bursts are indicated with arrows in Figure 
\ref{fig:asm}. The burst described in this work is similar in its
properties to the  other observed bursts as reported by
\citet{vanderspek05,barbier05,galloway05,galloway06}.
From the observed {\it INTEGRAL} burst properties and  mass
accretion rates  inferred  from  
the persistent luminosity, the present theory predicts that all these 
bursts are pure helium burning \citep[e.g.][]{s04}. For helium 
flashes, the fuel burns rapidly, since there are no slow weak 
interactions, and the local Eddington limit is often exceeded. These 
conditions lead to PRE bursts with a duration, set mostly by the time 
it takes the heat to escape, of the order of 5--10 s, as 
observed. In the framework of the 
thermonuclear-flash models \citep[e.g.,][]{lewin95} the burst 
duration, $\tau < 20$ s, and the ratio of 
 observed persistent flux to  net peak flux $\gamma\approx0.02$ indicate
a hydrogen-poor burst.

Because there were no other bursts observed during the {\em INTEGRAL}
observation, the burst recurrence time, $\Delta t_{rec}$,  must be at
least one day. We can compute the ratio of the total energy 
emitted in the persistent flux to that emitted in  the burst 
$$
\alpha= \frac{F_{pers}}{f_b} \Delta t_{rec} = \frac{\gamma}{\tau} \Delta t_{rec}  = 
 \frac{0.021}{18.2} \Delta t_{rec}  > 100, 
$$
which is consistent with pure helium bursts \citep[see e.g.][]{lewin95}. 
Taking  the burst total energy release $E_{b}=5\times 10^{39}$ erg   
(derived from the fluence $f_b$ reported in Sec. \ref{sec:spec_burst}) 
and  He burning efficiency of  
$\epsilon_{\rm He}\approx 1.7$ MeV/nucleon $\approx 1.6\times10^{18}$
erg g$^{-1}$, we can estimate the amount of fuel burned during the burst 
$E_{b}/\epsilon_{\rm He}\sim3.1\times10^{21}$ g. 
For the mean mass accretion rate of 2 per cent of the Eddington
through the entire outburst, a burst recurrence time of 2.2 days is expected.
This is consistent with the observed frequency of the bursts (see
Fig. 1 and note  that not all bursts have been detected).

\section{Conclusions} 
\label{sec:conclusions} 

We have found that the source spectrum is similar to other accreting
X-ray millisecond pulsars having a high plasma temperature around 30
keV and a Thomson optical depth $\sim2$. 
This source differs from other MSP in requiring thermal 
soft X-ray emission with nearly double the temperature. 
From our spectral fits we infer that this emission is not
likely be produced in a multi-temperature accretion 
disk but  more likely arises from thermal emission at the NS
surface. One might expect that the
lack of pulsations could be due to a particularly high optical depth,
but our spectral fits rule out this possibility.

The reason for the lack of coherent pulsations in the persistent emission
from LMXBs is still an open question. Different explanations have
been put forward to explain this phenomenon, including models which
invoke gravitational lensing, electron scattering, or weak
surface magnetic fields due to magnetic screening \citep[][and
  references therein]{w88,brainerd87,titarchuk02,cumming01}. The high
accretion rate inferred for \hete\ relative to the other known MSP transients
suggests that we may be observing the evolution of the NS's magnetic
field due to magnetic screening in this source \citep{cumming01}.
The  transition of \hete\ from an X-ray
millisecond pulsar to a persistent LMXB could indicate that
there is a population of suppressed  X-ray millisecond pulsars among the
non-pulsating LMXBs. Detailed observations of this source at the epoch
of pulsation suppression can help to solving the long-standing issue of
missing pulsations in persistent LMXB emission.

\begin{acknowledgements} 
MF acknowledges the French Space Agency  (CNES) for financial support.  
JP acknowledges the Academy of Finland grants 102181 and 110792.  
EWB is supported by Marie Curie Incoming European Fellowship contract
MIF1-CT-2005-008762 within the 6th European Community Framework
Programme.

\end{acknowledgements}

\end{document}